\documentclass[aps,twocolumn,prl,superscriptaddress]{revtex4}

\usepackage{graphicx}
\usepackage{dcolumn}
\begin{document}

% You should use BibTeX and revtex.bst for references
\bibliographystyle{revtex}
% marks overfull lines with blackboxes
%\draft

%\preprint{}

\title{Magnetic domain formation in itinerant metamagnets}
% Optional argument for running titles on pages
%\title[]{}



\author{B. Binz}
\email{binzb@berkeley.edu}
\affiliation{Theoretische Physik, ETH Zurich CH-8093 Zurich,
  Switzerland}
\affiliation{Department of Physics, University of Fribourg, Ch.~du Mus\'ee 3,
  CH-1700 Fribourg, Switzerland}
\affiliation{Department of Physics, University of California, %366 Le Conte \# 7300,
 Berkeley, CA 94720, %- 7300, 
 USA}
\author{H.~B.~Braun}
\affiliation{Theoretische Physik, ETH Zurich CH-8093 Zurich,
 Switzerland}
\affiliation{UCD School of Physics, University College Dublin, Dublin 4, Ireland}
\author{T.~M.~Rice}
\affiliation{Theoretische Physik, ETH Zurich CH-8093 Zurich,
 Switzerland}
\author{M. Sigrist}
\affiliation{Theoretische Physik, ETH Zurich CH-8093 Zurich,
 Switzerland}


\date{November 14, 2005}

\def\up{\uparrow}
\def\down{\downarrow}
\newcommand{\eref}[1]{(\ref{#1})}

\newcommand{\ud}{\mathrm{d}}

\renewcommand{\v}[1]{{\bf #1}}

\newcommand{\be}{\begin{equation}}
\newcommand{\ee}{\end{equation}}

\begin{abstract}
We examine the effects of long-range dipolar forces on 
metamagnetic  transitions and  generalize the theory of 
Condon domains to the case of an itinerant electron system undergoing 
a first-order metamagnetic transition. We demonstrate that within a finite range of the applied field, dipolar interactions induce  a spatial modulation of the magnetization in the form of  stripes or bubbles. Our findings are consistent with recent observations in the bilayer ruthenate Sr$_3$Ru$_2$O$_7$.
\end{abstract}

\maketitle

Itinerant electron systems exhibiting metamagnetism, quantum criticality and
non-Fermi-liquid behavior continue to attract widespread experimental \cite{grigera01,perry04,grigera04} and theoretical \cite{millis02,binz04,green04} interest
with  the bilayer ruthenate Sr$_3$Ru$_2$O$_7$ as
the most prominent example. Particularly surprising was the recent discovery of a new phase with increased resistivity as the quantum critical point is approached \cite{perry04,grigera04}.  Different theories have been proposed for the nature of this new phase  \cite{green04}.
In this Letter, we show that dipolar magnetostatic forces in an itinerant metamagnet lead to 
domain formation in a small yet finite region of
 the temperature field phase diagram. 
 These domains differ in the magnitude rather than the direction
 of the magnetization. Magnetic domain formation of this type was first proposed by Condon 
 to explain de Haas - van Alphen measurements in Be \cite{condon66}. Much later, Condon domains,  which are  formed and destroyed in each de Haas - van Alphen cycle, were observed directly
using muon-spin-rotation ($\mu$SR) spectroscopy \cite{solt96-02} and recently 
the domain pattern was observed  in Ag by micro Hall probes \cite{kramer05}.
 The theory has been further refined in the context of 
 de Haas - van Alphen effect \cite{itsovsky94-03},  but to our knowledge, 
 magnetic domain formation has not yet been considered in the 
 context of metamagnetism.

\begin{figure} 
\hfill\includegraphics[scale=0.35]{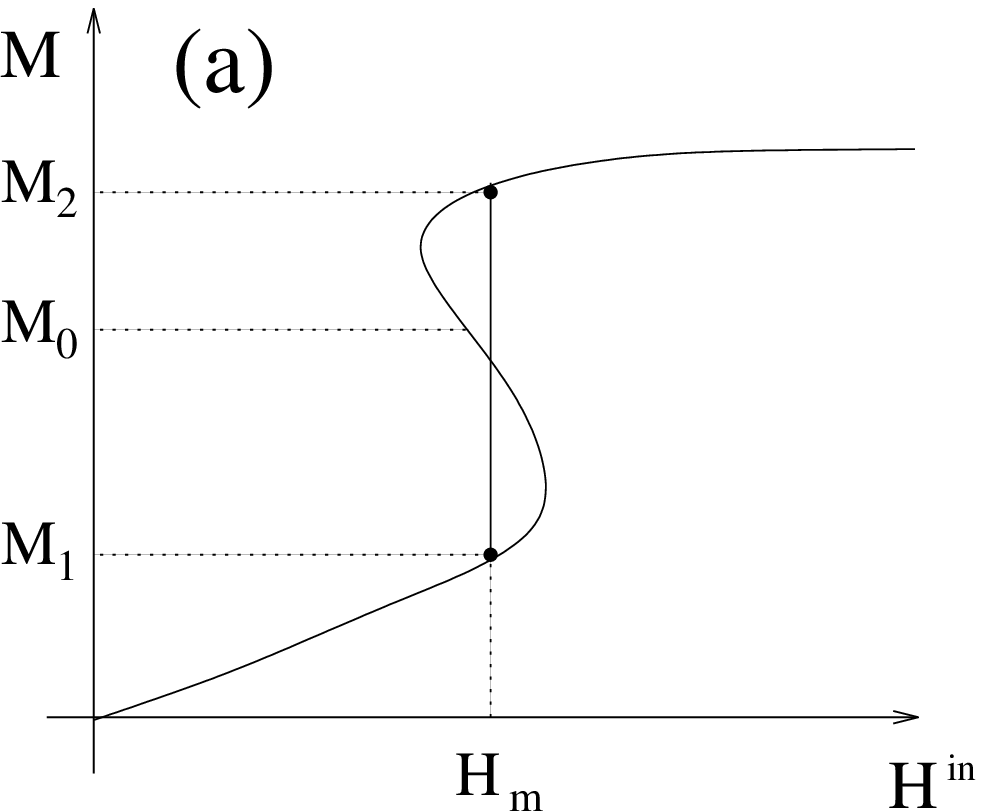}
\hfill\includegraphics[scale=0.35]{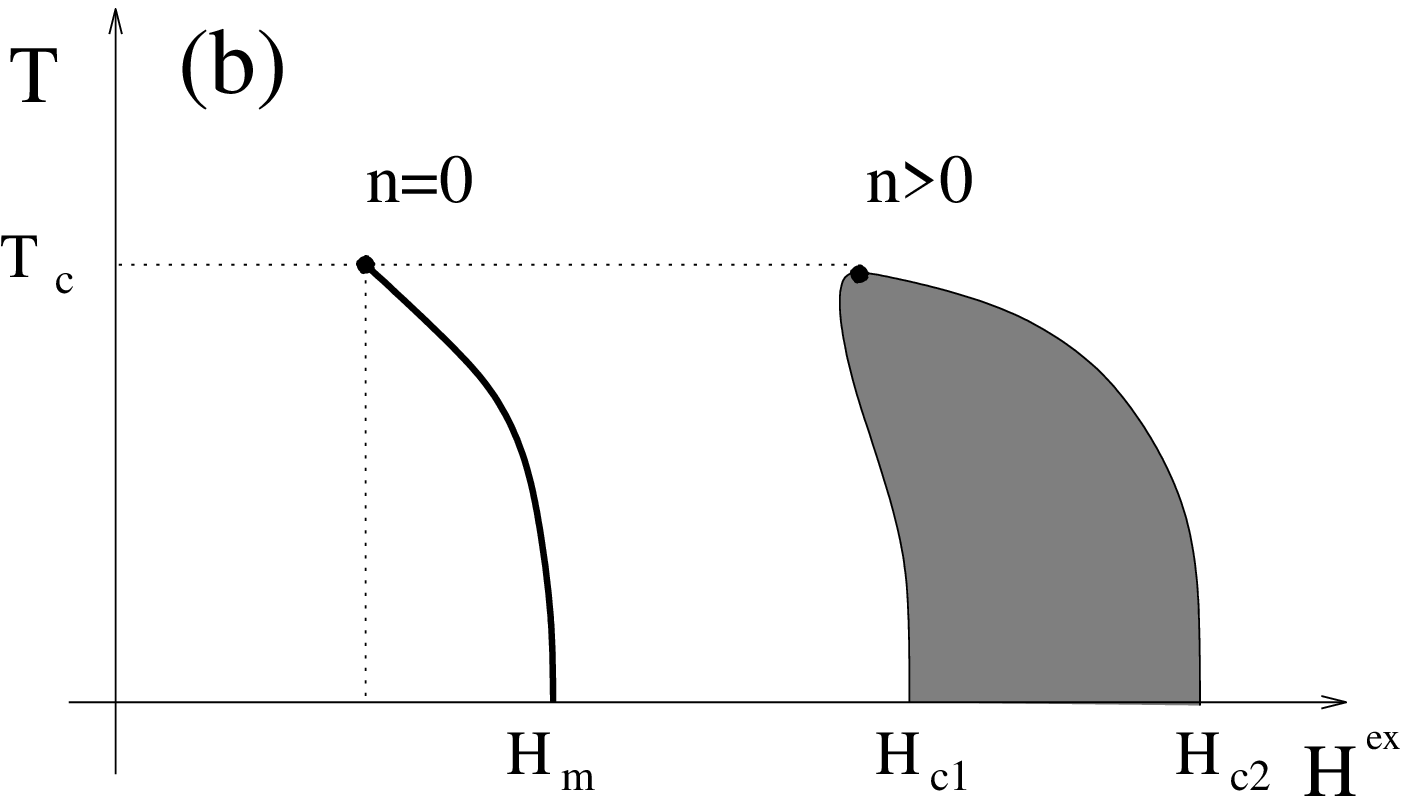}\hfill
\caption{(a) Magnetization as a function of the local internal magnetic field $H^{\rm in}$ with 
a  jump at field $H_{m}$. 
(b) Schematic phase diagram in the $H$-$T$ plane. For $n=0$, there is a line of first-order
  transitions terminating at a critical end-point. We assume that $H_m$ is a decreasing function of temperature, reflecting the fact that entropy increases with increasing magnetization in itinerant metamagnets \cite{binz04}. The grey area
 indicates the region of phase separation for  $n>0$.
\label{figs}} 
\end{figure}

We first consider briefly the standard phenomenological theory of itinerant metamagnetism
starting from the free energy per unit volume $f(M,T)$ as a function of magnetization (produced by spin polarization of a metallic electron band) and temperature. The thermodynamic equation of state is $\mu_0 H^{\rm in}=\partial_M f$, where $\mu_0$ is the 
vacuum permeability and $H^{\rm in}$ is the magnetic field inside the sample. The differential susceptibility is $\chi=dM/dH^{\rm in}$, or equivalently $\mu_0\chi^{-1}=\partial_M^2f$. By definition, metamagnetism occurs if the susceptibility has  a maximum at a finite field, i.e.~if $\partial_M^2f$ has a minimum at $M=M_0$. In cases where $\partial_M^2f$ becomes negative, a thermodynamic instability is produced where  the magnetization jumps abruptly between two values $M_1$ and $M_2$ at which $H^{\rm in}=H_{m}$ [Fig.~\ref{figs}(a)], determined by a Maxwell construction: $H_{m}=\partial_Mf|_{M_1}=\partial_Mf|_{M_2}=(f(M_2)-f(M_1))/(M_2-M_1)$. The location of the jump defines a line of first-order transitions in the $H$-$T$ plane, which ends at a critical endpoint [Fig.~\ref{figs}(b)]. A quantum critical endpoint occurs if $T_c$ is suppressed to zero as a function of additional parameters. 

This standard picture must be supplemented 
%extended 
to include 
magnetic dipolar interactions. We demonstrate this by translating an old argument by Condon \cite{condon66} into the context of metamagnetism. The internal magnetic field
 is a combination of the external applied field $H^{\rm ex}$ and 
 the field created by the magnetized sample
$ H^{\rm in}\approx H^{\rm ex}-n M$,
where $n$ is the demagnetization factor depending on sample shape and field orientation ($n=0$ for a needle-like sample oriented parallel to the field and $n=1$ for a thin film oriented perpendicular to the field). Thermodynamic 
stability is lost if
$H^{\rm in}$ reaches $H_{m}$: from the low (high) field side, this occurs when $H^{\rm ex}=H_{c1}=H_{m}+n M_1$,
 ($H^{\rm ex}=H_{c2}=H_{m}+n M_2$). Instead of a single first-order 
 transition, there is a finite region, $H_{c1}< H^{\rm ex} < H_{c2}$ where the uniformly magnetized state is unstable [Fig.~\ref{figs}(b)]. In this region, as we will show, magnetization forms domains  with  $M=M_1$ or $M_2$. 

A complete theory  is based on  the free energy functional for a spatially 
varying magnetization  $M({\bf r})$  parallel to the $z$-axis \cite{perpfluct},
\be
F=\int\!d^3r\,\left[f(M)+K\left(\nabla M\right)^2\right]+E_d,
\label{F}
\ee
where $K$ is a parameter controlling the stiffness, and
\be
E_d=\frac{\mu_0}{8\pi}\int\!d^3r\,\int\!d^3r'\,M(\v r)\,\Lambda(\v r-\v r')\,M(\v r'), \label{Ed_3D}
\ee
with  $\Lambda(\v r)=(r^2-3z^2)/r^5$ is  the dipolar magnetostatic 
energy. The equilibrium magnetization profile $M(\v r)$, which 
minimizes $F- \mu_0\int\!d^3r\,  H^{\rm ex} M$, 
is  determined  by a competition between
domain wall (DW) and dipolar energies. 
Below, we discuss its properties in the different regions of the phase diagram. 

Close to the critical temperature, one may expand $f$ in powers of  
$m(\v r)=M(\v r)-M_0$, where $M_0$ is the local minimum of 
$\partial^2_Mf$. Hence,
\be
f(M)=f(M_0)+\mu_0H_0 m + t m^2+u m^4+O(m^5),
\ee
where $\mu_0H_0=\partial_Mf|_{M_0}$. All the parameters $M_0, H_0$ and $t$ are temperature-dependent. Apart from the linear term which shifts the critical point to finite fields, we obtain the standard theory of an Ising ferromagnet 
with dipolar interactions \cite{garel82}. There is no third-order term by construction, but other odd terms such as $m^5$ are permitted. For simplicity, we
now consider an infinite film of thickness $D$. For this geometry with the field direction perpendicular to the slab, the demagnetization factor is $n=1$. If the magnetization is independent of the $z$-coordinate, the quadratic part of $F$ in powers of $m$ is %$D \int\!\!\frac{d^2\!q}{(2\pi)^2} \,\omega_{q}|m_{\v q}|^2$, 
$D \int\!{d^2\!q} \,\omega_{q}|m_{\v q}|^2/(2\pi)^2$, 
where 
$m_{\v q}=\int\!d^2r\,e^{i\v q \cdot  \v r}m(\v r) $ and 
$\omega_{q}=t+Kq^2+{\mu_0}(1-e^{-qD})/(2qD)$. 
We assume that the  sample is sufficiently thick that $e^{-qD}$ may be neglected. No matter how small the dipolar interaction may be in relation to the spin stiffness, the instability always occurs at a finite wave vector $q_0= \left(2l^2D\right)^{-1/3}$, where $l^2=2K/\mu_0$, giving sinusoidal modulations of the magnetization with wavelength $d=2\pi/q_0$. On following the line $H^{\rm ex}=H_0+M_0$ in the $H^{\rm ex}$-$T$ phase-diagram from high temperatures, there is a second-order transition from the uniform to the modulated phase as soon as the instability condition $t<-\frac32\mu_0\left(\frac{l}{2D}\right)^{2/3}$ is met. For values of the external field different from $H_0+M_0$, both  bubble- and stripe-modulated phases may be stabilized within mean-field theory and the transitions are weakly first-order  \cite{garel82}. In each case, the modulated phase consists of a smooth, sinusoidal spin-density wave on top of the uniform magnetization background.

\begin{figure*}
\includegraphics[scale=0.45]{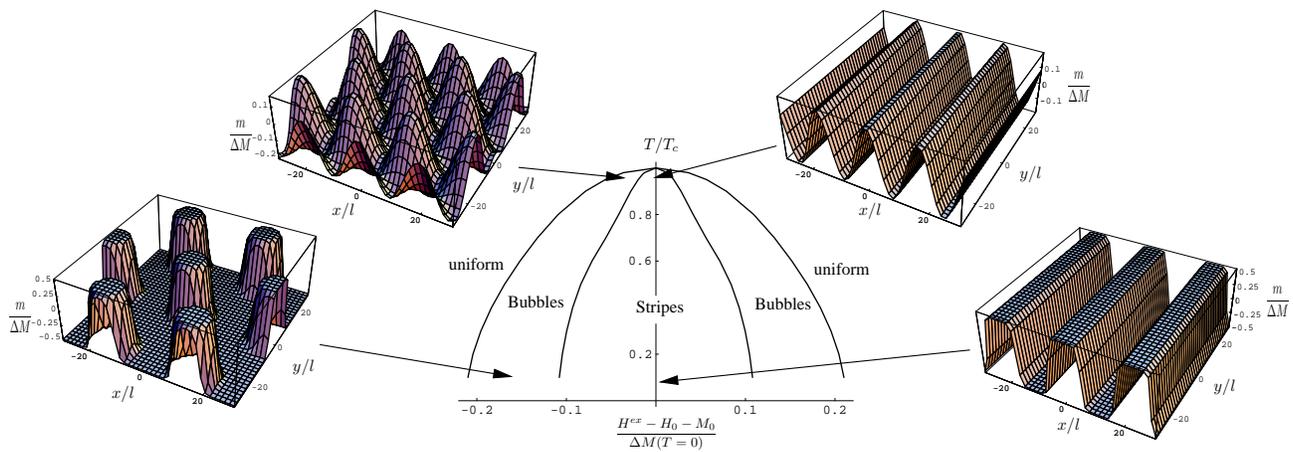}
\caption{(Color online) Mean-field phase diagram as a function of external magnetic field $H^{\rm ex}$ and temperature $T$ for the model of Eq.~\eref{flog} within the Ansatz of \eref{As}, \eref{Ab}. The parameters  chosen were $\gamma=100\mu_0$, $\delta=-8\mu_0$ and $D=20\,l$, where $l=(2K/\mu_0)^{1/2}$. The insets show the magnetization profile $m(\v 
r)=M(\v r)-M_0$ at four different locations. The transitions are first-order at low temperature, weakly first-order in the critical regime $T\alt T_c$ and second-order at $T=T_c$.
\label{pdiag}}
\end{figure*}

When  $T\ll T_c$, the double-well structure of $g(M)=f(M)-\mu_0H_{m} M$ 
becomes more pronounced, such that sinusoidal magnetization oscillations 
are no longer favored. The sample splits into domains 
of constant magnetization $M_1$ and $M_2$, separated by relatively 
sharp DWs. The determination of 
the domain structures then involves two independent problems. 
 The first is the internal structure of the DW, which depends on the potential $g(M)$ and the stiffness $K$, and the second  is 
 the global domain pattern, which depends solely on the 
 long-ranged dipolar interaction and the characteristic length $\xi=\sigma/(\mu_0\Delta M^2)$, where  $\Delta M=M_2-M_1$ is the difference of magnetization of the two phases and $\sigma$ is the  DW energy per unit surface. The first problem is easily solved. The equilibrium magnetization profile of a single, flat DW satisfies 
$dM/dx=\sqrt{(g(M)-g(M_1))/K}$,  which leads to the 
following
 characteristic DW width and energy:
\be
\lambda=\lim_{\epsilon\to0}\frac1{\ln (1/\epsilon)}\int_{M_1+\epsilon \Delta M}^{M_2-\epsilon \Delta M}\!dM\,\frac{\sqrt{K}}{\sqrt{g(M)-g(M_1)}},\label{lambda}
\ee
%The DW energy is 
\be
\sigma=2\sqrt{K}\int_{M_1}^{M_2}\!dM\,\sqrt{g(M)-g(M_1)}.\label{sigma}
\ee
The second problem has been studied in great detail in the context of uniaxial (Ising) ferromagnets in the shape of thin platelets 
far below $T_c$ \cite{cape71,kaczer70}. In an intermediate range of the ratio $D/\xi$, the phase diagram as a function of magnetic field in the sharp-wall limit is very similar to that in the critical regime:  A stripe phase in the center and a bubble phase at the border of the coexistence region separated by
first-order transitions. The typical domain size is of order $d\sim(\xi D)^{1/2}$, and differs
from the wavelength $2\pi/q_0$ in the critical regime. The domain structure is more complicated for thick samples ($\xi\ll D$), where an increasing number of wedge-shaped domains are formed close to the sample surface while the  
bulk domain size is typically $d\sim(\xi D^2)^{1/3}$ \cite{kaczer70}.  

In order to interpolate between these two regimes,  the critical regime close to $T_c$ and the sharp-wall regime at low temperature, we propose the following Ansatz, which depends on three variational parameters: $\lambda$, $R$, and $d$. For periodic stripe domains, 
\be
M_{\lambda,R,d}(x)=M_1+\frac{\pi\lambda\Delta M}{d}\sum_{q_\nu}\frac{\sin(q_\nu R)\cos(q_\nu x)}{\sinh\frac{\pi q_\nu \lambda}2}, \label{As}
\ee 
where $q_\nu =2\pi \nu/d$. For a triangular lattice of cylindrical bubble domains,
\be
M_{\lambda,R,d}(\v r)=M_1+\frac{2\pi^2\lambda R\Delta M}{\sqrt{3}d^2}\sum_{\v Q_{\nu\mu}}\frac{J_1(Q_{\nu\mu}R)\cos(\v Q_{\nu \mu}\cdot\v r)}{\sinh\frac{\pi Q_{\nu\mu}\lambda}2},\label{Ab} 
\ee
where $J_1$ is a Bessel function of the first kind and 
$\v Q_{\nu\mu}=\frac{2\pi}d(\nu,\frac{(2\mu+\nu)}{\sqrt{3}})$. 
In both cases $d$ is the period of the structure and $\mu,\nu$ are integers.
The interpretation of the parameters $R$ and $\lambda$ depends on the regime. If $\lambda\ll R,d$,
the Ansatz consists of domains of radius $R$ with magnetization $M_2$ in a  sea of  $M_1$ and $\lambda$ corresponds to the DW thickness. For $\lambda\agt d$, the Ansatz yields sinusoidal spin-density waves because higher Fourier modes are suppressed exponentially. The mean value around which the magnetization oscillates is controlled by $R$ and the amplitude 
by $\lambda$. Examples of stripe and bubble configurations in both regimes are illustrated in
Fig.~\ref{pdiag}. 

After these general considerations, we now consider a model for a 
specific itinerant metamagnet, namely a 2D electron system 
whose Fermi level is close to a logarithmic van Hove singularity  (VHS).
For this situation, the free energy $f(M,T)$ was 
calculated within Hartree-Fock theory in Ref.~\cite{binz04}.  Here, we present an effective model which captures the primary physical features. At the Hartree-Fock level, $\mu_0\chi^{-1}=\partial_M^2f\propto\rho^{-1}_{\uparrow}+\rho^{-1}_{\downarrow}-2I$, where $\rho_{\uparrow}$ and  $\rho_{\downarrow}$ are the thermally broadened densities of states at the Fermi level of up- and down-spin electrons and $I$ is the on-site Coulomb energy. At zero temperature, a singularity of the form $\partial_M^2f\sim1/\ln |M-M_0|$ occurs, where $M_0$ is the magnetization at which one of the two electron species has its Fermi level exactly at the VHS. The main effect of a small, finite temperature is to provide a cutoff to this logarithm. These observations motivate the following effective free energy [Eq.~\eref{F}]:
\be
\Delta f(M,T)=\left(
{\delta-\gamma}/{\ln
\left[{\left({m}/{m_0}\right)^2+{T}/{T_0}}\right]}\right)m^2,\label{flog}
\ee
where $m_0$ and $T_0$ are positive parameters,  $m=M-M_0$ and $\Delta f=f-\left.f\right|_{M=M_0}-\mu_0H_0 m$.  
Since $\Delta f$ is an even function of $m$, it follows that $H_m=H_0$ for this model. 
The model of Eq. ~\eref{flog} is meaningful only for 
 $m\ll m_0$ and $T\ll T_0$. 
The parameter $\gamma$ is of the order $vW/\mu_B^2$, where $v$ is the 
volume of a unit cell  and $W^{-1}$ is 
the weight of the VHS.
The parameter $\delta$ is proportional to the doping away from the quantum critical endpoint in the mean-field phase diagram neglecting $E_d$ \cite{binz04}. Dipolar interactions shift the  quantum critical endpoint to $\delta_c=-\frac32\mu_0\left({l}/{2D}\right)^{2/3}$. For  $\delta>\delta_c$, there is a metamagnetic crossover and for $\delta<\delta_c$ a first-order metamagnetic transition with a coexistence region below 
$
T_c=T_0 \exp\left\{{\gamma}/({\delta-\delta_c})\right\}.
$
At  zero temperature and assuming sharp DWs $\lambda\ll d$, we %calculate using the Maxwell construction, Eq.~\eref{flog} and Eqs.~\eref{lambda} - \eref{sigma}:
find using the Maxwell construction and Eqs.~(\ref{lambda},\ref{sigma},\ref{flog}):
\begin{eqnarray}
\Delta M &=&2m_0\exp\left\{({\gamma+\sqrt{\gamma(\gamma-4\delta)}} ) /({4\delta})\right\},\\
\lambda/l&=&\left[0.98+O\left(\delta/\gamma\right)\right]\sqrt{\mu_0\gamma}/(-\delta),\\
\xi/l&=&\left[0.42+O\left(\delta/\gamma\right)\right] (-\delta)/\sqrt{\mu_0\gamma}.
\end{eqnarray}
The crossover between the critical and the sharp-wall regime occurs at $1-T/T_c\approx(2l/D)^{2/3}\mu_0\gamma/\delta^2$. 

\begin{figure}
\includegraphics[scale=0.5]{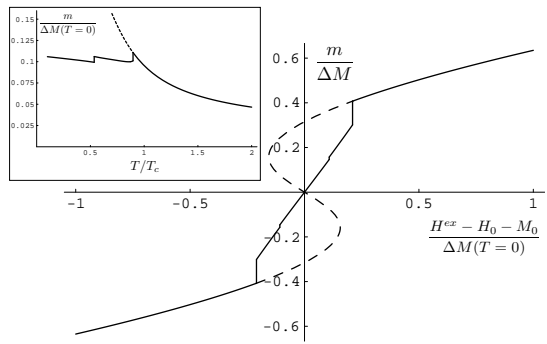}
\caption{Mean magnetization $m=M-M_0$ as a function of the external 
field at $T=0.1\,T_c$. Model and parameters are as in Fig.~\ref{pdiag}. The dashed line indicates the (unstable) solution if the magnetization were uniform. The solid line was obtained 
with the Ansatz of \eref{As}, \eref{Ab}. Inset:
Temperature dependence of the magnetization along the line $H-H_{0}-M_0 = 0.08\,\Delta M(T=0)$. 
\label{m_h_model}}
\end{figure}

%Fig. \ref{pdiag} shows  a $H^{\rm ex}$-$T$ phase diagram for the model of Eq.~\eref{flog}, which was obtained numerically by using the Ansatz of \eref{As} and \eref{Ab}. 
The entire  $H^{\rm ex}$-$T$ phase diagram for the model of Eq.~\eref{flog} is shown in Fig. \ref{pdiag}. It was obtained numerically by using the Ansatz of \eref{As} and \eref{Ab} and minimizing $F- \mu_0\int\!d^3r\,  H^{\rm ex} M$.
Figure \ref{m_h_model} shows the mean magnetization as a function of $H^{\rm ex}$ at low temperature. The abrupt increase of the magnetization both on entry and on exit from the domain phase are clearly visible. Much smaller increases appear also at the transitions between bubble and stripe phases. The energy difference between the bubble and stripe solutions is extremely small, so that the precise domain structure will be susceptible to many details, e.g.   lattice structure, impurities, and  sample shape. The inset of Fig. \ref{m_h_model} shows the temperature dependence of the magnetization 
for constant $H^{\rm ex}$.  Note the increase of $m$  with decreasing temperature in the uniform phase due to the diverging susceptibility at $H^{\rm in}=H_0$. Upon entering the domain phase, the magnetization increase
is cut off since the internal field $H^{\rm in}$ is locked to the nearly temperature independent value of $H_{m}$. A small magnetization jump occurs at both first-order phase boundaries.

The existence of magnetic domains  is known to increase the electrical resistivity due to DW scattering. 
In uniaxial ferromagnets, the DW resistivity shows a sharp increase upon entering the domain phase by 
tuning the magnetic field at low temperature \cite{ravelosona99}. The resistivity in the direction perpendicular to oriented DWs is higher than in the parallel direction \cite{viret00} and even shows a non-metallic behavior \cite{feigenson03}. In contrast to ferromagnetic domains, Condon domains differ by the 
magnitude rather than the direction of their magnetization. This implies that properties such as the Fermi velocity vary across  Condon DWs leading to strong scattering of electrons, especially if the Fermi level is near a VHS, where 
electronic properties depend strongly on the band filling. 

Our results are particularly relevant in view of the recent observation of a new phase in Sr$_3$Ru$_2$O$_7$ with a field oriented along the $c$-direction \cite{perry04,grigera04}. The qualitative similarities between the phase diagram (Figs.~\ref{figs} and \ref{pdiag}) and the phase diagram %established in 
of Ref.~\cite{grigera04} are striking. Also, if metamagnetism is caused by a logarithmic VHS, %in the density of states, 
then the domain phase is very sensitive to sample purity, because the VHS itself is very sensitive to impurities. The authors of Ref.~\cite{grigera04} give a list of five key experimental facts which place  rather tight constraints on any  theory for the new phase. We  note that all five points are compatible with magnetic domain formation. (i) DW scattering  leads to an enhanced resistivity, (ii) the electrons are itinerant throughout  the phase diagram, (iii) the mean magnetization increases both on entry and on exit from the domain phase as a function of field (cf.~Fig.~\ref{m_h_model}), (iv) a
rise of the mean magnetization is cut off as the domain phase is entered from high-temperature (cf. 
inset of Fig.~\ref{m_h_model}), and (v) the domain phase exists only in a narrow range of applied fields. From the observed field range $\mu_0(H_{c2}-H_{c1})\approx 0.2T$ \cite{grigera04} and 
the volume $78 \mbox{\AA}^3$  per Ru,
 we estimate $n\Delta M\approx 1.3$ $\mu_B/{\rm Ru}$. While this number appears rather large, 
 it is not in error by orders of magnitude. Finally we note that we have restricted ourselves to 
 the simplest case of a field along the $c$-axis. 
 Other orientations  require corrections beyond the scope of this letter.
$\mu$SR, NMR spectroscopy and micro Hall probe technics could allow a direct detection 
of domain formation.  For (quasi-) static domains the difference in  local fields will lead to a 
characteristic splitting  in the resonance lines  as long 
as the corresponding dynamics is slow compared to the experimental time scales. 

In conclusion, we predict that an itinerant 
electron system 
 in general 
cannot undergo a first-order phase transition 
without breaking up into magnetic domains
within a finite range of applied fields. While we cannot rule out other mechanisms
explaining the unusual feature of the low temperature 
phase in Sr$_3$Ru$_2$O$_7$,
our results show that a detailed experimental analysis is  required  
to decide if mechanisms other than Condon domains underlie the unusual behavior.

\acknowledgments
B.B. would like to thank C. Capan, E. Demler, A.G. Green, Z.Q. Mao, B. Normand, C. Pfleiderer, A. Scholl and A. Vishwanath for helpful and stimulating discussions. This work was supported by the NCCR MaNEP,  the Swiss National Science Foundation and the Science Foundation Ireland.

\end{document}